# Extracting chemical food safety hazards from the scientific literature automatically using large language models

Neris Özen[a], Wenjuan Mu[a], Esther D. van Asselt[a], Leonieke M. van den Bulk[a,*]

[a] *Wageningen Food Safety Research, Wageningen University & Research, Wageningen, The Netherlands*

\* *Corresponding author, e-mail address: leonieke.vandenbulk@wur.nl*

*Abstract*

The number of scientific articles published in the domain of food safety has consistently been increasing over the last few decades. It has therefore become unfeasible for food safety experts to read all relevant literature related to food safety and the occurrence of hazards in the food chain. However, it is important that food safety experts are aware of the newest findings and can access this information in an easy and concise way. In this study, an approach is presented to automate the extraction of chemical hazards from the scientific literature through large language models. The large language model was used out-of-the-box and applied on scientific abstracts; no extra training of the models or a large computing cluster was required. Three different styles of prompting the model were tested to assess which was the most optimal for the task at hand. The prompts were optimized with two validation foods (leafy greens and shellfish) and the final performance of the best prompt was evaluated using three test foods (dairy, maize and salmon). The specific wording of the prompt was found to have a considerable effect on the results. A prompt breaking the task down into smaller steps performed best overall. This prompt reached an average accuracy of 93% and contained many chemical contaminants already included in food monitoring programs, validating the successful retrieval of relevant hazards for the food safety domain. The results showcase how valuable large language models can be for the task of automatic information extraction from the scientific literature.

1. **Introduction**

It is crucial that food safety experts are aware of the newest research concerning the occurrence of hazards in the food chain to ensure our food can be as safe as possible. Because of the growing interest in food safety over the last decades, the number of scientific articles related to the domain has consistently been increasing (Luo et al., 2022). However, this has introduced a challenge: Hundreds of publications are currently written on food safety each year, and it has become unfeasible for food safety experts to read everything relevant for their field. This makes it difficult to keep track of all new findings and potential risks related to the food safety domain. To increase awareness of new hazards so that timely action can be taken to safeguard our food, it is important that information reaches the food safety experts in an easy and concise way. Automating the extraction of relevant information from scientific literature can be a promising asset to save researchers' valuable time.

Natural language processing (NLP) can be adopted for this task, which has seen a rise in development in the past few years. NLP is the branch of Artificial Intelligence (AI) that focuses on understanding text and is used for text translation (Luong et al., 2015; Wu et al., 2016; Zhu et al., 2023), summarization (Y. Liu & Lapata, 2019; Zhang et al., 2020), part-of-speech tagging (Chiche & Yitagesu, 2022; Wang et al., 2015), text classification (Kowsari et al., 2019; Minaee et al., 2021), and text generation (Brown et al., 2020; Devlin et al., 2019; Lewis et al., 2019). NLP can also be used for information extraction, where relevant information from unstructured text is collected automatically without the need for a domain expert to manually go through the text (Hong et al., 2021; Zaman et al., 2020). Early research in NLP focused on the application of rule-based methods or n-gram models (K. Min & Wilson, 2006; Nadkarni et al., 2011), but these approaches have long been overtaken by better performing neural network models (Khurana et al., 2023; B. Min et al., 2023). Moreover, in recent years there has been a shift towards using pre-trained neural network models instead of training these models from the ground up on task-specific datasets (Howard & Ruder, 2018; Peters et al., 2018; Vaswani et al., 2017). Pre-trained NLP models have been trained on enormous amounts of texts with the goal of capturing the underlying patterns and structures that are present in human language. In contrast to task-specific datasets, these texts do not have to be labelled, and can instead be trained using self-supervised learning (X. Liu et al., 2021). This makes pre-training much less labor and cost intensive, allowing the possibility to train these models on vast amounts of data. When models are trained on large datasets and use advanced neural network architectures, they become capable of both understanding and generating realistic and human-like text. These models are referred to as large language models (LLMs) and they have quickly acquired popularity in the AI domain for the past few years, due to their capability of performing various tasks at state-of-the-art levels (Khurana et al., 2023; B. Min et al., 2023).

LLMs can either be finetuned on a task-specific dataset to specifically be calibrated to perform a single task or they can be used by giving them instructions in natural language without any additional training, also called prompting. Radford et al. (2019) and Brown et al. (2020) pioneered in this domain by demonstrating that scaling up language models could eliminate the need to further finetune them for specific tasks with labeled datasets. Their LLMs were some of the first models which only required a textual instruction to carry out a particular linguistic task, at a performance comparable to that of state-of-the-art finetuned models. When ChatGPT was released, Qin et al. (2023) explored the performance of ChatGPT across a wide range of tasks to investigate if it was a "general-purpose" model for NLP. They found that although ChatGPT was not yet a true generalist model, it still showed remarkable capabilities and obtained good scores across the full range of tasks, which included arithmetic reasoning, question answering, summarization and information extraction. The specific prompt given to the model, however, is very important since the quality of the prompt can have a major effect on the quality of the answer (P. Liu et al., 2023). The formulation of optimal prompts has evolved into an emerging scientific field, known as prompt engineering. Various strategies can be applied when generating prompts to enhance the performance of the LLM, such as setting clear goals for the task, providing distinct subtasks, giving examples of expected answers and specifically asking for the reasoning behind the answer (P. Liu et al., 2023; Santu & Feng, 2023).

Automatic information extraction can be a powerful tool to obtain knowledge and insights from any domain, drastically reducing the need for manual work. Since LLMs are trained to understand the context behind a text, they are currently the start-of-the-art when it comes to information extraction (Xu et al., 2023). LLMs have been applied widely across different scientific domains for the purpose of information extraction, and have especially been developed within the biomedical domain. LLMs have already been successfully applied in that domain for the case of medical information extraction from clinical notes (Agrawal et al., 2022), finding adverse effects of drugs in medical reports (Gu et al., 2023), summarizing the effect of food on the absorption of drugs from drug application reviews (Shi et al., 2023), and identifying the number of participants from scientific publications (Paroiu et al., 2023). There has also been a growing interest in applying LLMs in solving the challenges of the food domain. They have been applied for the generation of healthy diet suggestions for people with food allergies (Niszczota & Rybicka, 2023), the extraction of food dish names from restaurant reviews and social media posts (Lin et al., 2023), and the creation of a question-answer system of food testing data (Qi et al., 2023).

In this study, we aimed to assess whether LLMs can automatically extract chemical food safety hazards from the scientific literature through prompting without any additional training of the models, making the approach easy to implement for non-AI experts. We used scientific abstracts as our input for the LLM, as these are readily available in literature databases, contain the most important findings and offer a concise text that can be processed well by language models. This approach can showcase how valuable large language models can be for the task of automatic information extraction from the scientific literature in the field of food safety.

## 2. Materials and Methods

An open-source LLM was applied for the extraction of chemical contaminants from scientific abstracts by prompting the model with textual instructions, and without introducing any extra training for the specific task at hand. The pipeline presented in Figure 1 shows the performed steps. To prompt the LLM with relevant scientific abstracts and correctly identify the chemical names in the responses of the LLM, data collection and preprocessing steps were required for both the set of abstracts and a chemical dictionary. After filtering the collected abstracts related to a specific food, the LLM is prompted to extract the mentioned chemical contaminants in the abstracts. The final steps aim to parse the responses of the LLM to generate the final list of chemical hazards for a food item and evaluate their correctness. Evaluation was performed for both a validation and test procedure. A validation procedure was used to identify the prompt that elicited the best performance from the LLM using leafy greens and shellfish as validation foods. This was followed by a test procedure to measure the definitive performance of the best-performing prompt using dairy, maize and salmon as test foods. The detailed procedures related to the steps in the pipeline are described in the subsections below.

### 2.1. Abstracts collection and preprocessing

The open-access literature repository Europe PMC was used to collect scientific abstracts that were of interest for this study. Relevant abstracts should include chemical contamination in food with a possible negative effect on human health. A search query was created based on two sets of search terms and each abstract was required to contain at least a word or phrase from each set of terms in its title, abstract or keywords. The first set of terms focused on words/phrases indicating presence of chemical hazards, whereas the second set of terms contained words/phrases focusing on an impact on public health. The search query was based on those used in systematic literature reviews previously performed to identify chemical contaminants in food (Banach et al., 2019; Kluche et al., 2020). See Appendix A for the final search query that was used.

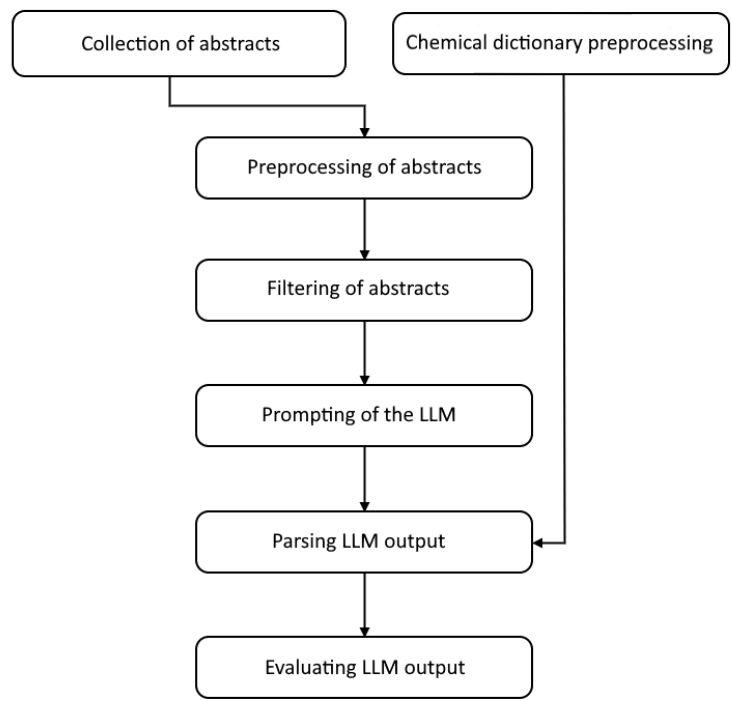

*Figure 1. The pipeline showing the performed steps to extract chemical hazards from scientific abstracts using an LLM.*

To obtain the abstracts and corresponding metadata, a request was made to the REST API service of Europe PMC with the created search query. All abstracts with a publication date up to April 2nd, 2023, were stored. The DOI, title and publication year of the original article were stored along with the abstracts.

The collected abstracts were cleaned by removing HTML tags, excessive spaces and copyright declarations. Empty, duplicate and abstracts shorter than 60 characters were excluded, in addition to abstracts correcting or amending previously published research.

## 2.2. Chemical dictionary preprocessing

Even though the LLM is asked to only retrieve chemical hazards, there is a possibility that it returns a different entity (e.g. a microbiological or physical hazard). Furthermore, chemicals often have many different synonyms that can be used. To link these different synonyms across the abstracts and to make sure that our output only contains chemical hazards, we applied a chemical dictionary to identify correct names of chemicals in the responses of the LLM. The dictionary chosen for this study was ChEBI (Chemical Entities of Biological Interest). ChEBI includes a controlled vocabulary where for each entry, a chemical name is provided alongside its unique ChEBI identifier, IUPAC name and its synonyms (Hastings et al., 2016). The dictionary contains over 170,000 unique chemicals. With chemical names and their synonyms provided alongside a persistent ChEBI identifier, different names or different writing conventions of the same chemical seen in the responses of LLM can be matched to a single identifier (e.g. both 'Cd' and 'cadmium' match to the same identifier CHEBI:28628).

The dictionary was preprocessed to provide the best applicability to our case. Some entries were removed from consideration because their names are too generic (e.g. "molecule", "solvent" or "vitamins") or don't represent chemical entities (e.g. "application", "voltage" or "alpha"). In addition, extra synonyms using different writing conventions for existing chemical names were added to the dictionary. This step was executed on chemical names which contain numbers alongside letters. For example, in case a chemical belongs to a specific isotope of an element (e.g. polonium-210) or a specific type of compound (e.g. aflatoxin b1), more synonyms were created by swapping the number and word components or adding or removing spaces or hyphens (e.g. generating 210-polonium or aflatoxin b-1).

The dictionary is also extended to contain all plural forms of the chemicals. The final preprocessed dictionary contains approximately 1.4 million chemical names.

## 2.3. Abstract filtering and LLM prompting

Abstracts were filtered to only retain abstracts relevant to a selected food. The collected abstracts from Europe PMC were filtered based on the presence of keywords defining the respective foods from the validation and test procedures. For the foods used for validation, abstracts for leafy greens were required to include "leafy green", "leafy greens", "leafy vegetable" or "leafy vegetables", whereas for shellfish, abstracts were required to include "shellfish". For the foods used to test the final performance, in the case of dairy "dairy" needed to be present in abstracts, for maize "maize" or "corn" was required and finally for salmon "salmon" needed to be present.

Once abstracts for the desired food were determined, an open-source LLM was used for information extraction. To ensure the reproducibility of our work, remove the need for a remote cluster with substantial computational power and with privacy concerns of LLM platforms in mind, we decided to employ an open-source LLM that is relatively small in terms of its parameters and can be run locally on a personal computer with a good GPU. The "Nous-Hermes-13B-GPTQ" LLM model was selected based on these criteria and downloaded from the Hugging Face platform[1], a platform where AI models are openly shared. Nous-Hermes LLM models are made available by Nous Research, a company focused on open-sourcing LLM models that are small enough to be run locally on people's own computers. The Nous-Hermes-13B-GPTQ model capitalizes on the Llama architecture and has been finetuned on instructions generated with synthetic GPT-4 outputs (Touvron et al., 2023). The model was able to rival GPT-3.5 in performance.

---

[1] https://huggingface.co/TheBloke/Nous-Hermes-13B-GPTQ

The hyperparameters in the LLM were chosen towards keeping randomness in its responses to a minimum and ensuring reproducibility as much as possible. To that end, hyperparameters were set as such: do_sample = False, num_beams = 1, repetition_penalty = 1.0, which forces the LLM, at each step of generating its response, to always choose the output with the highest probability, creating a consistent answer.

The LLM was prompted by providing each of the filtered abstracts one by one, with the instruction to extract the chemical hazards and the food(s) they are present in from that abstract. Since the performance of an LLM in a task can vary significantly depending on the exact prompt, three different styles of prompts were drafted and assessed for the accuracy of their outputs on the validation foods. Specific wording in the different prompts were incrementally updated according to their performance. The three styles of prompts can be characterized as follows:

1. "Simple prompt": This is written fully in natural language. The task the LLM is expected to execute and the required format of the LLM output are expressed in two sentences. In light of earlier experiments with the LLM and the observation of its pitfalls, another paragraph warning the LLM against potential pitfalls was added. At the very end, the abstract of interest is provided in triple backticks.

2. "Step by step prompt": As recommended by Fulford & Ng (2023), the task is broken down into 5 smaller tasks and each smaller task is expressed in enumerated steps to facilitate a reasoning process and to prevent the LLM from giving rash answers. With the same motivation, the model is asked to print the outcome of each step in a specific format. To make sure that this prompt does not have a disadvantage compared to the "simple prompt" concerning the pitfalls the LLM is prone to, the same paragraph of warning is provided in this prompt as well. Like in the "simple prompt", the abstract is provided in triple backticks for the model to execute the task on.

3. "Pseudo code prompt": This style of prompting is inspired by the work of Mishra et al. (2023), who defend that pseudo-code instructions are less susceptible to ambiguities and find that pseudo-code instructions beat their counterparts in natural language over a variety of NLP tasks. The task is introduced in the format of a Python function where the smaller tasks constituting the task are now expressed as a line of code in the function, accompanied by natural language description of the smaller task as comment. Docstrings are used to repeat the natural language descriptions of all smaller tasks and include the warnings against the common pitfalls. After the Python function is defined, it is called with the abstract as input.

Full textual instructions for each prompt are provided in Figure 2, 3 and 4.

### 2.4. LLM output parsing and evaluation

The LLM was instructed to extract the chemical hazards and their corresponding foods from each abstract in a dictionary format, where a food item is linked to a list of its hazards (e.g. {'Chinese cabbage': ['cadmium', 'chromium'], 'wheat': ['deoxynivalenol', 'arsenic'], 'shellfish': ['saxitoxin']}). If there are no chemical hazards present in an abstract, the model was asked to output an empty dictionary. Chemical hazards were only accepted as answers if the corresponding food item extracted by the LLM contained one of the keywords used to filter the abstracts as specified in section 2.3.

The responses of the LLM were mostly consistent in providing the requested output format as shown above, however, initial evaluation showed that they sometimes deviated from the required format (e.g. the hazards were not provided in a list or the hazards were embedded inside another dictionary). The responses in deviant formats did, however, tend to contain correct chemical hazards. Therefore, when parsing the responses, the most common deviations from the desired output format were accommodated.

```
┌─────────────────── Simple prompt ───────────────────┐
│                                                     │
│  Extract foods and chemicals that are mentioned to be a food safety hazard
│  for one of the foods, to contaminate one of the foods or to have the
│  potential to pose risk for human health via consumption of one of the foods
│  in the text delimited by triple backticks.
│  Provide the output in dictionary format with each different food as a
│  separate key and the chemicals that are expressed to be hazardous,
│  contaminant or to have the potential to pose risk for human health via
│  consumption of the food as the value of the key.
│
│  I want to warn you against some pitfalls. First, make sure that you bring
│  each chemical name that is mentioned to be a contaminant, hazard,
│  potentially harmful for human health via food consumption, especially make
│  sure not to skip the specific compound names. Another thing is if chemicals
│  are mentioned both with their names and abbreviations, make sure to return
│  the full name of the chemical instead of its abbreviation. Next warning -
│  only provide foods and chemicals that are mentioned in the text provided,
│  do not return any food or chemical that is not mentioned in the text. Also,
│  do not try to provide more specific foods or chemicals if the foods or
│  chemicals in the text are only mentioned in their general category. Another
│  thing - refrain from providing irrelevant noun phrases or sentences in
│  values just because they contain chemical names, limit the values of your
│  dictionary to the names of relevant chemicals. Also, return an empty
│  dictionary if you do not identify any chemical in food as contaminant,
│  hazardous, potentially harmful for human health through consumption of
│  food. Finally, limit your answer to the dictionary, no other explanation or
│  justification is necessary.
│
│  ```{ABSTRACT}```
│
└─────────────────────────────────────────────────────┘
```

*Figure 2. The full prompt text for the "simple prompt".*

The parsed output was compared against the preprocessed chemical names and synonyms from ChEBI. This helped with filtering out the non-chemical names in the list, as the LLM returned some adjectives (e.g. "hazardous") or some microbiological hazards in a number of responses. In addition, some of the responses can contain an abbreviation of a chemical instead of its full name. In this situation, the abbreviations are replaced by the full chemical name by tracing back the abbreviation in the original abstract. The found chemical names were subsequently matched to their ChEBI identifiers. This way, the same chemical extracted in varying writing conventions from different responses were not counted double.

```
Step by step prompt

Your task is to perform the following actions:
1. Identify the chemicals mentioned in the text below provided between
triple backticks and collect them in a list
2. Identify the foods mentioned in the text below provided between triple
backticks and collect them in a list
3. Create all combinations of foods and chemicals as tuples and collect the
tuples in a list
4. Go over each food-chemical combination in the list created at step 3 and
look whether the chemical is mentioned to be a food safety hazard for that
food, to contaminate that food or to have the potential to pose risk for
human health via consumption of that food. Store each food-chemical
pair where chemical is said to be hazardous, contaminant or to have the
potential to pose risk for human health via consumption of the food, in a
dictionary where foods are keys and chemicals that are expressed to be
hazardous, contaminant or to have the potential to pose risk for human
health via consumption of the food are values.
5. Once you go over every food-chemical pair, return the dictionary you
obtained.

I want to warn you against some pitfalls. First, make sure that you bring
each chemical name that is mentioned to be a contaminant, hazard,
potentially harmful for human health via food consumption, especially make
sure not to skip the specific compound names. Another thing is if chemicals
are mentioned both with their names and abbreviations, make sure to return
the full name of the chemical instead of its abbreviation. Next warning -
only provide foods and chemicals that are mentioned in the text provided,
do not return any food or chemical that is not mentioned in the text. Also,
do not try to provide more specific foods or chemicals if the foods or
chemicals in the text are only mentioned in their general category. Another
thing - refrain from providing irrelevant noun phrases or sentences in
values just because they contain chemical names, limit the values of your
dictionary to the names of relevant chemicals. Also, return an empty
dictionary if you do not identify any chemical in food as contaminant,
hazardous, potentially harmful for human health through consumption of
food. Finally, limit your answer to the dictionary, no other explanation or
justification is necessary.

Use the following format:
Chemicals: <chemicals you identified in the text below between triple
backticks>
Foods: <foods you identified in the text below between triple backticks>
Dictionary: <dictionary storing food-chemical pairs where foods are keys
and chemicals expressed to be hazardous, contaminant or have the potential
to pose risk for human health via consumption of the food item are values>

```{ABSTRACT}```
```

*Figure 3. The full prompt text for the "step by step prompt".*

A list of ChEBI identifiers of chemicals identified as hazardous and the corresponding abstract DOI's was created to facilitate the evaluation procedure. Evaluation of the results for the validation foods involved

```
                              Pseudo code prompt

    def identify_safety_hazards(text: str) -> dict:

        """Identify the chemical and food items mentioned in the provided abstract. For each combination
        of chemical substances and foods, look whether the chemical substance is mentioned to be a food
        safety hazard for that food, to contaminate that food, has possibility to pose risk for that food
        in future or has the potential to pose risk for human health via food chain. Keep food-chemical
        substance pairs where chemical substance is said to be hazardous, contaminant, potential future
        risk for the food or to pose risk for human health, in a dictionary where food items are keys and
        the hazardous, contaminant, potentially risky or potentially harmful for human health chemical
        substances are values. Once you go over every food-chemical pair, return the dictionary you
        obtained. I want to warn you against some pitfalls. First, make sure you bring each chemical
        substance name that is mentioned to be a contaminant, hazard, potential risk or harmful for human
        health, especially make sure not to skip the specific compound names. Another thing is if chemical
        substances are mentioned both with their names and abbreviations, make sure return the full name
        of the chemical instead of its abbreviation. Next warning - only provide foods and chemical
        substances that are mentioned in the text provided, do not return any food or chemical substance
        that is not mentioned in the text. Also, do not try to provide more specific foods or chemical
        substances if the foods or chemicals in the text are only mentioned in their general category.
        Another thing - refrain from providing irrelevant noun phrases or sentences in values just because
        they contain chemical substance names, limit the values of your dictionary to the names of relevant
        chemical substances. Also, return an empty dictionary if you do not identify any chemical substance
        in food as contaminant, hazardous, potentially risky or harmful for human health. Finally, limit
        your answer to the dictionary, no other explanation or justification is necessary."""

        #Create an empty dictionary
        chemical_hazards_per_food = {{}}

        #Identify the chemical items mentioned in the provided text and collect them in a list
        chemical_list = identify_chemicals_in_text(text)

        #Identify the foods items mentioned in the provided text and collect them in a list
        food_list = identify_foods_in_text(text)

        #Create all combinations of food and chemical items as tuples and collect the tuples in a list
        food_chemical_combinations = [(food, chemical) for food in food_list for chemical in chemical_list]

        #Go over each food-chemical combination and look whether the chemical is mentioned
        # to be a food safety hazard for that food, to contaminate that food or have
        # the potential to pose risk for human health via consumption of that food
        for food, chemical in food_chemical_combinations:

            #Store food-chemical pairs where chemical is said to be hazardous, contaminant
            # or to have the potential to pose risk for human health via consumption of the food,
            # in a dictionary where foods are keys and chemicals that are expressed to be hazardous,
            # contaminant or to have the potential to pose risk for human health via consumption of
            # the food are values
            if chemical_is_hazardous_food(food, chemical, text) and food in chemical_hazards_per_food:
                chemical_hazards_per_food[food].append(chemical)

            elif chemical_is_hazardous_food(food, chemical, text) and food not in chemical_hazards_per_food:
                chemical_hazards_per_food[food] = [chemical]

            else:
                continue

        return chemical_hazards_per_food

    >>> identify_safety_hazards('{ABSTRACT}')
```

*Figure 4. The full prompt text for the "pseudo code prompt".*

manually checking the DOI's provided for each chemical entry. It was checked that the abstracts indeed expressed that that chemical, or the wider contaminant group, contaminates that food. At the end of the validation stage, the best performing prompt was chosen for ultimate assessment in the test cases. For each test food, the results returned by the LLM were evaluated by a domain expert following the same approach as in the validation stage. In both procedures, the performance was measured by the number of correct chemical hazards extracted from scientific abstracts for a specific food.

3. Results

In total 101,727 unique abstracts were retrieved from Europe PMC with the aim to identify chemical contamination in food with a possible negative effect on human health. These abstracts were filtered to only retain abstracts related to the selected validation and test foods. The total number of abstracts relevant for the validation foods were 411 for leafy greens and 1235 for shellfish. For the test foods 1403, 1318 and 353 abstracts were filtered as relevant for dairy, maize and salmon respectively. Figure 5 shows the number of abstracts per year for each of the foods, clearly showing an upward trend in the number of published abstracts on this topic. Note that the year 2023 was omitted due to abstracts only having been collected until April.

Table 1 provides the accuracy of the retrieved chemical hazards returned by the LLM for each of the validation and test foods for the tested prompts. Results are reported in two ways: number of correct responses divided by the total number of responses and the percentage that comes from this division (indicated in parentheses). The "simple prompt" had the worst performance among the three for the validation foods, with a performance of 89.2% for shellfish and 64.5% for leafy greens. The "Step by step prompt" had a higher accuracy than the "pseudo code prompt" in leafy greens with 100% compared to 93.8%. For shellfish the "pseudo code prompt" had marginally higher accuracy with 92.9% than the "step by step prompt" with 92.4%. It should be noted that the "step by step prompt" returned a substantially higher number of correct chemical hazards compared to the "pseudo code prompt", resulting in 5 more hazards in leafy greens and 37 more in shellfish. Considering both the accuracy and returned number of chemical hazards, "step by step prompt" was chosen as the best performing prompt to be evaluated on the test foods. The performance of the "step by step prompt" on the test foods was highest for dairy with 98.7%, followed by 92.0% for maize and 88.9% for salmon. The number of correctly retrieved chemical hazards was 75, 69 and 48 for dairy, maize and salmon respectively. A full list of the found chemicals for the test foods can be found in appendix B.

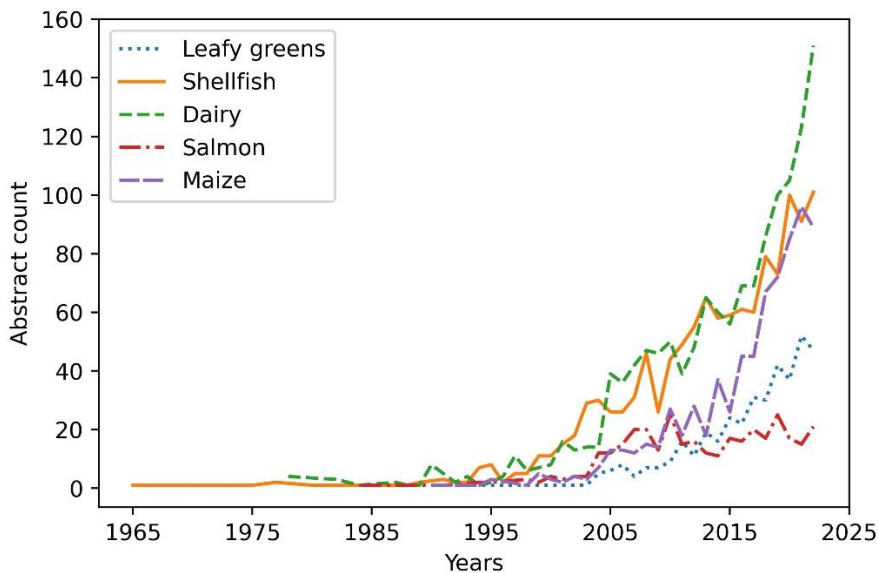

*Figure 5. Counts of abstracts per year retrieved from Europe PMC for each of the validation and test foods.*

Correctly retrieved chemical hazards for dairy included veterinary drugs (e.g. corticosteroids, chloramphenicol and tetracycline), heavy metals (e.g. arsenic, cadmium and uranium), mycotoxins (e.g. aflatoxin B1+M1, deoxynivalenol and fumonisins), organic contaminants (e.g. bisphenol A+F, perfluorooctanoic acid and hexabromocyclododecane), pesticides (e.g. hexachlorobenzene, pentachlorophenol and deltamethrin) and biogenic amines (e.g. histamine, tyramine and cadaverine).

| Prompt style | Validation foods | | Test foods | | |
|---|---|---|---|---|---|
| | Leafy greens | Shellfish | Dairy | Maize | Salmon |
| Simple prompt | 20/31 (64.5%) | 91/102 (89.2%) | -/- (-%) | -/- (-%) | -/- (-%) |
| Step by step prompt | **21/21 (100%)** | **73/79 (92.4%)** | **75/76 (98.7%)** | **69/75 (92.0%)** | **48/54 (88.9%)** |
| Pseudo code prompt | 15/16 (93.8%) | 39/42 (92.9%) | -/- (-%) | -/- (-%) | -/- (-%) |

*Table 1. Accuracy of the returned chemical hazards by the LLM for both the validation and test foods. All prompts were applied on the abstracts of the validation foods and the best prompt, which was the "Step by step prompt" was applied on the abstracts of the test foods.*

For maize, the chemical hazards included mostly mycotoxins (e.g. ochratoxin A+B, beauvericin and sterigmatocystin), heavy metals (e.g. cadmium, chromium and mercury), pesticides (e.g. carbaryl, thifluzamide and chlorpyrifos) but also polycyclic aromatic hydrocarbons (e.g. pyrene) and pyrrolizidine alkaloids. The chemical hazards classified as correct for salmon included heavy metals (e.g. methylmercury, zinc and selenium), pesticides (e.g. emamectin benzoate, hexachlorobenzene, chlordane), veterinary drugs (e.g. medetomidine, ofloxacin and quinolones), polycyclic aromatic hydrocarbons (e.g. naphtalene, fluorene, phenanthrene), aromatic hydrocarbons (ethylbenzene, toluene and xylene), organotins (dibutyltin) and persistent organic contaminants (e.g. dioxins, furans and polychlorinated biphenyls).

Table 2 provides the chemical hazards retrieved for the test foods that were marked as incorrect by the domain expert, per food. Even though some of the chemicals are seen as food safety hazards, they were classified as incorrect since they were either not mentioned in the abstract or not in connection to the food studied. For dairy, carvacrol is the sole incorrect chemical. Our domain expert explains this to be a beneficial compound in oregano oil, which is used as a feed additive. The expert has marked arsenite, sulfonate, nitrate, aflatoxin M1, ellagic acid and fenpyrazamine as incorrect for maize. Arsenite was not present in the abstract the LLM collected the answer from and was therefore marked as incorrect. Sulfonate was incorrect because the abstract actually mentioned "6:2 fluorotelomer sulfonate" instead, but this was not present as an entity in ChEBI, while sulfonate was. Nitrate and aflatoxin M1 were marked as incorrect as they did not occur in maize in their abstracts, but in soil and milk respectively. Meanwhile, ellagic acid is a beneficial bioactive compound and therefore was not classified as a chemical hazard. Lastly, even though it is a hazardous chemical, fenpyrazamine was mentioned to be dangerous for corn salad, which is a type of leafy green, and is not a hazard for corn/maize. For salmon the expert marked aminobenzoic acid, chitosan, trimethylamine, mirex, polycyclic aromatic hydrocarbons and astaxanthin as incorrect. Aminobenzoic acid, mirex and polycyclic aromatic hydrocarbons were marked incorrect as they were not present in their corresponding abstracts. Chitosan and astaxanthin are described as unharmful chemicals added for prolonging the shelf life and enhancing the color of salmon respectively. Trimethylamine is the main spoilage product in salmon, but is not considered a food safety hazard.

| Dairy | Maize | Salmon |
|---|---|---|
| Carvacrol | Arsenite | Aminobenzoic acid |
| | Sulfonate | Chitosan |
| | Nitrate | Trimethylamine |
| | Aflatoxin M1 | Mirex |
| | Ellagic acid | Polycyclic aromatic hydrocarbons |
| | Fenpyrazamine | Astazanthin |

*Table 2. Chemical hazards identified as incorrect in the LLM responses by the domain expert for each test food.*

## 4. Discussion

This study showed that chemical contaminants in foods can be extracted from scientific abstracts using an open-source LLM. This was done by providing textual instructions and without finetuning for the specific task at hand, making it easy to implement and recreate. Three differently styled prompts were evaluated by measuring accuracy of the responses of the LLM for a specific food item (i.e. leafy greens and shellfish) to determine the best performing prompt. The "step by step prompt" was established as the best prompt and was further evaluated for dairy, maize and salmon. Based on the evaluation of the domain expert, the performance for dairy, maize and salmon are very promising with accuracies of 98.7%, 92.4% and 88.9% respectively.

Overall, this research set out to perform an information extraction task. Interestingly, some of the incorrect answers the LLM provided for the test cases could have been made by anyone without enough specific food safety domain knowledge. One such example is the misclassification of aflatoxin M1 in maize. The LLM extracted this result from an abstract that discussed aflatoxin M1 levels in milk samples, where they also tested maize-based cow feed for aflatoxin contamination (Yunus et al., 2020). The LLM classified aflatoxin M1 as an aflatoxin also present in the feed, but aflatoxins in maize are only converted into aflatoxin M1 after being consumed by cows. Nonetheless, domain knowledge is required to know that feed can't contain aflatoxin M1 and the abstract does not specify this. Another example is trimethylamine in salmon. The LLM extracts trimethylamine from an abstract discussing it as a spoilage product (Jaaskelainen et al., 2018). The formation in salmon is undesirable, and a reader with no specific knowledge might not have been able to assess whether trimethylamine is only a food quality or also a food safety issue. Similarly, for both carvacrol and astaxanthin it was stated in the abstract that they are an irritant to skin and eyes, but that as an additive they do not raise any safety concerns (Bampidis et al., 2022; Rychen et al., 2017). Since the prompts refer to retrieving contaminants with risks to human health, this is an understandable mistake. In light of these examples, it is obvious that this task cannot only be considered a reading comprehension task. Therefore, it is always important to check the findings of the LLM with a domain expert.

The most interesting behavior observed in the output of the LLM was its provision of specific hazardous chemicals in its responses when only the larger contaminant group is mentioned in the abstract (e.g. retrieving specific veterinary drugs when veterinary drugs are only mentioned as a group), despite it being explicitly instructed to only retrieve contaminants explicitly mentioned in the abstract. Initially, this explicit warning was included to prevent randomly crafted, incorrect answers. However, it is noted that during the validation and test procedures, the specific chemicals returned from abstracts only mentioning their wider group of hazards, are indeed hazardous for the food item of interest. This points to the LLM having a broader implicit knowledge of the food safety domain which it uses to give more specific results than just a group of contaminants. Therefore, these results are not penalized and are also considered correct. The LLM for example returns antimicrobials such as ampicillin and ciprofloxacin for dairy, which are indeed correct hazards, based on an abstract solely referring to the "antimicrobials" in water collected from milking parlors (Veiga-Gomez et al., 2017). Furthermore, in the case of salmon, organic contaminants such as lindane, heptachlor and chlordane, known to be contaminants in salmon, were returned by the LLM from an abstract discussing the levels of organic contaminants in an area known for its salmon farming industry in New Zealand (Niu et al., 2023).

Many chemical hazards currently monitored in food safety programs for the tested foods were present among the parsed output of the LLM, validating the approach by its successful retrieval of relevant hazards for the food safety domain. However, some relevant chemical contaminants present in the abstracts were not among the output. This was mostly caused by the LLM providing either an incorrect output format, causing the output to not be parsed correctly, or the contaminant being linked to a subcategory of the food instead. For example, a contaminant could be retrieved for milk, which should then also be linked to dairy. Future research could try to solve this problem by connecting food with a food ontology, such that foods can automatically be linked to their broader categories which creates a more complete picture of the contaminants present in food (Munir & Anjum, 2018).

Similarly, finetuning an LLM for the specific task at hand is a conventional, yet another encouraging avenue for future research. Although it is more time-intensive and computationally costly to finetune an LLM, it will very likely reduce the deviations from the requested output format significantly, as it will have been trained with that format in mind. This makes parsing the output easier and could lead to the identification of more chemical hazards. Furthermore, the probability of a correct chemical contaminant not being extracted from an abstract will presumably become even lower. Developments to address the computational and memory costs associated with finetuning of LLMs are actively being worked on and have led to novel and promising approaches such as Low-Rank Adaptation (LoRA) (Hu et al., 2021).

The prompt styles experimented with in this study included three common techniques of prompting. However, in the ever-expanding literature on prompt engineering and LLMs many more styles and techniques do exist. It would have been too time consuming to try each possible approach presented in the literature to identify the best performing style. In contrast, the study by Yang et al. (2023) proposed an iterative approach in which LLMs are instructed with a "meta-prompt". In this prompt the LLM is asked to create a prompt for a specific task itself, and the newly created prompt is improved step by step by the LLM in line with the performance evaluated on a small dataset. They showed the prompts created by the LLM outperform those created by humans. Adopting this approach would be another promising direction for future research.

This research laid the foundation for a tool that food safety experts can use to get an overview of hazardous chemicals for any food. With automation of data collection and LLM prompting, such a tool can be automatically updated with new scientific literature and therefore be ensured to remain up-to-date. Links to the abstracts from which the LLM extracted the chemical hazards can be provided, so that the findings can be checked in case a result is deemed to be unexpected or unrealistic. Additionally, the frequency with which a chemical has been discussed in the literature for a specific food could be supplied. If a hazard has not been discussed often before, this could be used by food safety experts as a potential indication of an emerging risk and assist them in taking action around such risks more swiftly.

5. Conclusion

In this study, the application of an LLM for the extraction of chemical hazards from the scientific literature for specific foods was demonstrated. It was shown that the LLM can successfully retrieve relevant chemical contaminants and makes few errors. Across a set of three different foods, the average correct response rate was 93%. Multiple styles of prompts were tested to assess which was most optimal for the task at hand. The specific wording and style of the prompt was found to have a considerable effect on the performance. It was concluded that a prompt breaking the task down into smaller steps performed best overall. Using this approach, the information collection from scientific literature on chemical hazards in food can be automated and save researchers' valuable time. By automatically updating the results with the newest literature and adding frequency on how often a contaminant has been mentioned for a specific food, the approach could provide a way of detecting emerging hazards so that timely action can be taken to improve food safety.

*Code and data availability*

The code and data used in this study can be found on

https://github.com/WFSRDataScience/LLMForChemicalFoodSafetyHazardExtraction.

*Acknowledgements*

This research received funding from the Dutch Ministry of Agriculture, Nature and Food Quality (LNV) via the Knowledge Base Program "Healthy and Safe food systems" (KB-37-002-036).

*Author contributions*

**Neris Özen:** Methodology, Software, Validation, Writing - original draft, Writing - review & editing.

**Wenjuan Mu:** Methodology, Supervision, Writing - review & editing. **Esther D. van Asselt:** Validation, Writing - review & editing. **Leonieke M. van den Bulk:** Conceptualization, Methodology, Project administration, Supervision, Writing - original draft, Writing - review & editing.

*Declaration of competing interest*

The authors declare no competing interests.

**Appendix A**

**Final search query for Europe PMC:**

In title, abstract, keywords:  'food contamination' OR 'chemical pollutant*' OR 'chemical hazard*' OR 'contamina*' OR 'toxin*' OR 'toxic substance*' OR 'toxic compound*' OR 'pollutant*', 'agricultural chemical*', 'chemical compound*' OR 'chemical substance*' OR 'residu*'

AND

In title, abstract, keywords:  'public health' OR 'haccp' OR 'consumer protection' OR 'consumer*' OR 'food safety' OR 'risk assessment*' OR 'risk analys*' OR 'hazard analys*' OR 'human health*' OR 'health impact' OR 'health risk*' OR 'bioaccumulation'

**Appendix B**

| Dairy | Maize | Salmon |
|---|---|---|
| deoxynivalenol | deoxynivalenol | polybrominated diphenyl ether |
| ciprofloxacin | zearalenone | ethylbenzene |
| norfloxacin | tetraniliprole | mercury |
| sulfamethazine | quizalofop-p-ethyl | chitosan |
| polychlorinated dibenzofurans | ochratoxin b | naphthalene |
| polybrominated biphenyls | culmorin | pcb 138 |
| hexabromocyclododecane | mefentrifluconazole | pcb 180 |
| polybrominated diphenyl ethers | 15-adon | benzene |
| spermine | 3-acetyl-deoxynivalenol | perfluoroalkyl substances |
| tyramine | atrazine | toluene |
| iodide | mercury|mercury | trimethylamine |
| naphthalene | fumonisin b3 | iron |
| spermidine | ergosterol | histamine |
| formaldehyde | fluoride | aminobenzoic acid |
| putrescine | nitrate\|nitrate | dibutyltin |
| penicillin | sterigmatocystin | quinolones |
| iodine | aflatoxins | dioxins |
| pentachlorophenol | acetochlor | furans |
| tylosin | aflatoxin b1 | xylene |
| chloramphenicol | mycotoxins | arsenic |
| cadaverine | arsenic | selenium |
| sterigmatocystin | glyphosate | cobalt |
| histamine | lead | lead |
| aflatoxin | tungsten | fluorene |
| cephalosporin | chromium | cadmium |
| dioxins | dioxin | copper |
| furans | acrylamide | phenanthrene |
| aflatoxin b1 | cadmium | cu2+ |
| mycotoxins | molybdenum | zinc |
| amoxicillin | copper | methylmercury |
| uranium | arsenate | lindane |
| arsenic | arsenite | fluoranthene |
| selenium | beauvericin | polycyclic aromatic hydrocarbons |
| selenomethionine | lithium | pcb 153 |
| monensin | zinc | pcb 52 |
| chlortetracycline | terbuthylazine | chlordane |
| lead | bifenthrin | heptachlor |
| tetracycline | sulfonate | heptachlor epoxide |
| melamine | pahs | mirex |
| 2,3,7,8-tetrachlorodibenzodioxine | carbaryl | anthracene |
| daidzein | carbofuran | dibenzofurans |
| fluorene | chlorpyrifos | pyrene |
| semicarbazide | fumonisin b1 | emamectin benzoate |
| cadmium | fumonisins | astaxanthin |
| phenanthrene | fumonisin b2 | pentachlorobenzene |
| ampicillin | methoxyfenozide | endosulfan |
| sulfadimethoxine | pyrene | medetomidine |
| acenaphthylene | kojic acid | polychlorinated biphenyls |
| bisphenol a | deltamethrin | hexachlorobenzene |
| carvacrol | paracetamol | enniatin b |
| bisphenol f | ellagic acid | ofloxacin |
| perfluorooctanoic acid | microcystins | ethoxyquin |
| enrofloxacin | aflatoxin b2 | toxaphene |
| coccidiostats | citrinin | persistent organic pollutants |
| polychlorinated dibenzodioxine | cd2+ | bde 47 |
| fumonisins | butenolide | |
| dibenzofurans | zinc oxide nanoparticles | |
| deltamethrin | fusaric acid | |
| trimethoprim | polyethylene | |
| diclofenac | gliotoxin | |
| perchlorate | trichothecenes | |
| corticosteroids | enniatins | |
| polychlorinated biphenyls | pyrrolizidine alkaloids | |
| gliotoxin | patulin | |
| hexachlorobenzene | nivalenol | |
| ibuprofen | ochratoxin a | |
| lincomycin | aflatoxin m1 | |
| patulin | pyraclostrobin | |
| flunixin meglumine | aflatoxin g2 | |
| persistent organic pollutants | aflatoxin g1 | |
| aflatoxin m1 | thifluzamide | |
| bde-153 | fusarin c | |
| bde-99 | fenpyrazamine | |
| bde-47 | dimethenamid-p | |
| bde-209 | t-2 toxin | |
| ptaquiloside | | |
| sulfamethoxazole | | |

*Table B.1. All chemical hazards identified by the LLM for each test food.*